# Gate-tunability of the superconducting state at the $EuO/KTaO_3$ (111) interface

Weiliang Qiao[1†], Yang Ma[1†], Jiaojie Yan[1†], Wenyu Xing[1], Yunyan Yao[1], Ranran Cai[1], Boning Li[1], Richen Xiong[1], X. C. Xie[1,2,3], Xi Lin[1,2,3], Wei Han[1]*

[1]International Center for Quantum Materials, School of Physics, Peking University, Beijing 100871, P. R. China

[2]CAS Center for Excellence in Topological Quantum Computation, University of Chinese Academy of Sciences, Beijing 100190, P. R. China

[3]Beijing Academy of Quantum Information Sciences, Beijing 100193, P. R. China

[†]These authors contributed equally to the work

*Correspondence to: weihan@pku.edu.cn

**Abstract:**

The recent discovery of superconducting interfaces in the $KTaO_3$ (111)-based heterostructures is intriguing, since a much higher superconducting critical temperature ($T_C$; ~ 2 K) is achieved compared to that in the $SrTiO_3$ heterostructures (~ 300 mK). In this paper, we report the superconducting properties of $EuO/KTaO_3$ (111) interface as a function of the interface carrier density ($n_s$). The maximum $T_C$ is observed to be ~ 2 K at $n_s$ ~ $1\times10^{14}$ cm$^{-2}$. In addition, we show that the critical current density and the upper critical magnetic field can be effectively tuned by the back gate voltage. Interestingly, the gate dependence of the upper critical magnetic field exhibits a trend opposite to that of $T_C$ in the underdoped region, suggesting a relatively larger Cooper pairing potential.

## I. INTRODUCTION

The observation of superconducting interfaces between two insulating oxide layers of $SrTiO_3$ and $LaAlO_3$ is fascinating for the field of superconductivity [1-4]. Since then, a lot of efforts has been spent to achieve high temperature superconducting oxide interfaces via the heterostructure engineering and ionic liquid gating, such as $LaTiO_3/SrTiO_3$ [5], $Al_2O_3/SrTiO_3$ [6], and $KTaO_3$ (100) [7], etc. Recently, the superconductivity in $KTaO_3$ (111)- and $KTaO_3$ (110)-based heterostructures has been discovered with a superconducting critical temperature ($T_C$) up to ~ 2.2 K [8-11]. This $T_C$ is much higher than those observed in $SrTiO_3$ (100)-, $SrTiO_3$ (111)-, and the $KTaO_3$ (100)-based heterostructures [1-4,7]. Subsequently, several interesting physical phenomena have been observed, including the anisotropic transport behavior with a possible "stripe"-like phase [8], and the quantum Griffiths transition [9] as well as the electric field control of superconductivity [11]. The pairing mechanism of the superconducting $KTaO_3$ (111) interfaces poses a theoretical challenge [12]. Experimentally, the systematical investigation of the carrier density ($n_s$) dependence of various superconducting properties will be necessary for understanding the paring mechanism of the $EuO/KTaO_3$ (111) interfaces.

In this paper, we report the superconducting properties of $EuO/KTaO_3$ (111) interfaces as a function of $n_s$ tuned by varying the EuO growth condition and by applying a back gate on the $KTaO_3$ crystals. The $T_C$ reaches its maximum of 2 K with $n_s \sim 1\times10^{14}$ cm$^{-2}$. As $n_s$ varies, the critical current density ($J_C$) exhibits a trend similar to that of $T_C$, while the upper critical magnetic field ($B_{C2}$) exhibits the opposite trend. The opposite trends between $B_{C2}$ and $T_C$ in the underdoped region are similar to the characteristics of high temperature superconducting cuprates, suggesting a relatively stronger Cooper pairing potential at lower $n_s$. Our results could pave the way for further understanding of the pairing mechanisms of superconductivity interfaces in the $KTaO_3$ (111) heterostructures.

## II. EXPERIMENTAL

The EuO thin films were grown on the $KTaO_3$ (111) single crystals via oxide molecular beam epitaxy (MBE-Komponenten GmbH; Octoplus 400) at 500 ºC. There were two steps of EuO thin film growth, including the ~ 1 nm Eu thin layer without oxygen and the following ~ 9 nm EuO layer with the oxygen partial pressure of ~$1\times10^{-9}$ mbar. After the EuO growth, a 5-nm MgO layer was deposited via *e*-beam evaporation *in situ* to protect the EuO films from degradation during subsequent measurements. The EuO/$KTaO_3$ Hall bar devices were fabricated using standard photolithography followed by a wet-etching step using diluted hydrochloric acid. The details of the growth conditions and device fabrication process can be found in our previous report [9]. A silver paste was used on the back side of the $KTaO_3$ crystals to apply the gate voltage ($V_g$) to tune $n_s$.

The electron transport properties of EuO/$KTaO_3$ heterostructures were measured via either van der Pauw or standard Hall geometries. For $T$ from 300 to 1.5 K, the electrical measurements were conducted in an Oxford Spectromag system with the dc technique ($I_{dc}$ ~ 100 μA). The low temperature measurements from $T$ = 4 to ~ 0.26 K were performed in a He-3 insert (HelioxVT) in an Oxford TeslatronPT system. In this system, the resistances *vs*. temperature and magnetic field were measured using a low frequency lock-in technique ($I_{ac}$ = 1 nA, $f$ = 17 Hz or 23 Hz), and the critical current measurements were conducted using a dc source Keithley 2400 and a multimeter Keithley 2000. All the carrier densities were obtained in the Oxford Spectromag system using Hall measurements at $T$ = 10 K. The back gate voltage was applied using a Keithley 2400. Gate cycling was performed prior to the gate-dependence measurement and the normal state $R_s$ shows little change. To be noted, $V_g$ was varied in the range from -200 to 100 V to prevent electric breakdown of the $KTaO_3$ single crystals under $V_g$ higher than 100 V.

## III. RESULTS AND DISCUSSION

Figure 1(a) shows the schematic of the superconducting interfaces between the insulating EuO films and single crystalline $KTaO_3$ (111) substrates. As described in the experimental section, the interface $n_s$ can be controlled by changing the growth time of EuO in the first step. Generally speaking, the longer growth time of Eu results in higher carrier density for EuO/$KTaO_3$ (111) interfaces. The blue, red, black and green curves in Fig. 1(b) show the sheet resistance ($R_s$) as a function of temperature ($T$) for four representative samples from #1 to #4. From the zero-resistance temperature, $T_C$ can be identified to be 1.87 K, 1.75 K, 1.33 K and 1.23K for samples #1, #2, #3 and #4, respectively. The results of the onset temperature and uniformity of superconductivity state are shown in the Supplemental Material [13].The corresponding interfaces $n_s$ are determined to be $1.03 \times 10^{14}$, $9.65 \times 10^{13}$, $7.97 \times 10^{13}$, and $7.74 \times 10^{13}$ $cm^{-2}$ via Hall measurement at $T = 10$ K. These results indicate the critical role of $n_s$ in $T_C$, which is consistent with the previous report on the superconducting EuO/$KTaO_3$ interface [8].

To further investigate the $T_C$ dependence on $n_s$, a back gate is applied to finely tune $n_s$ of these as-grown samples, as schematically shown in Fig. 2(a) inset. Firstly, the gate voltage dependence of the sample #1 is studied, and three representative $R_s$-$T$ curves are shown in Fig. 2(a). Clearly, as $V_g$ varies from -150 to 50 V, the superconducting transitions are almost identical to each other. Figure 2(b) summarizes the results of $T_C$ vs. $V_g$. Despite the relatively large modulation of $n_s$ (measured at $T = 10$ K) as $V_g$ varies (inset of Fig. 2(b)), $T_C$ exhibits little variation. Besides, the $R_s$- $T$ curves of sample #4 with lower $T_C$ are shown in the Fig. 2(c) measured under $V_g$ from -200 to 100 V. Clearly, the $T_C$ under negative $V_g$ is significantly smaller than the $T_C$ under positive $V_g$. The $V_g$ dependences of $T_C$ and $n_s$ (measured at $T$ = 10 K) are summarized in Fig. 2(d). In general, $T_C$ and $n_s$ show a similar trend as $V_g$ varies between -200 and 50 V. For $V_g$ > 50 V, $T_C$ shows a slight decrease as $n_s$ increases from 7.8 × $10^{13}$ $cm^{-2}$ to 8.0 × $10^{13}$ $cm^{-2}$. This discrepancy needs future studies.

Figure 3 shows the summary results of $T_C$ vs. $n_s$ measured on 8 representative samples, where the solid symbols represent the results obtained from the as-grown samples, and hollow circles represent the results obtained under $V_g$. As $n_s$ increases, $T_C$ first increases, and then saturates with $n_s$ in the range from ~ 9.5 × $10^{13}$ to ~ 1.04 × $10^{14}$ $cm^{-2}$. The results of $n_s$ higher than ~ 1.04 × $10^{14}$ $cm^{-2}$ cannot be obtained despite a large effort by optimizing the growth conditions and applying $V_g$. Nevertheless, the saturation of $T_C$ may indicate the superconducting dome feature of the EuO/$KTaO_3$ interfaces, as illustrated by the dashed line in Fig. 3. This superconducting dome as a function of carrier density is similar to those observed in other two-dimensional superconductors, such as $LaAlO_3$/$SrTiO_3$ and $MoS_2$ [14-17].

Next, we investigate the gate tuning critical current density ($J_C$) of the superconducting EuO/$KTaO_3$ (111) interfaces. Figure 4(a) shows three representative current-voltage ($I$-$V$) curves measured on the sample #4 at $T$ = 0.26 ± 0.01 K. The critical current ($I_C$) is obtained to be ~ 2.6, ~ 10.6, and ~ 21.2 μA for $V_g$ = -200, -100, and 100 V, respectively. At $V_g$ = -100 and 100 V, hysteresis of $I$-$V$ curves around $I_C$ are observed, which could be attributed to the current heating effect of the superconducting interfaces [18]. At $V_g$ = -200 V, $I_C$ decreases significantly compared to that at $V_g$ = 100 V, and the hysteresis is no longer observable. Based on the width of the Hall bar device ($W$ = 100 μm), $J_C$ is calculated based on $J_C = I_C / W$, and the results of $J_C$ as a function of $V_g$ are shown in Fig. 4(b). The relationship of $J_C$ vs. $V_g$ follows a trend similar to that of $T_C$ vs. $V_g$, which is consistent with the previous report at the $LaAlO_3/KTaO_3$ (111) interface [11]. For sample #1, the critical current density ($J_C$) exhibits little variation as $V_g$ varies (the data is not shown), which agrees well with the results of $T_C$ vs. $V_g$ (Fig. 2(b)).

In the following, we discuss the gate dependence of $B_{C2}$. In the previous study [8], different $B_{C2}$ were observed with $B$ perpendicular and parallel to the sample plane, which demonstrates the two-dimensional feature of the superconducting $EuO/KTaO_3$ (111) interface. In the current work, we focus on the $B_{C2}$ that is perpendicular to the sample plane to investigate the Cooper pairing strength. Figure 5(a) shows the $R_s$ vs. $B$ curves measured at $T = 0.26$ K under various $V_g$ from -100 to 100 V. Clearly, compared to the results at $V_g$ = -100 V (red curve), $B_{C2}$ at $V_g$ = 100 V (blue curve) decreases significantly.

To quantitatively investigate the relationship between $B_{C2}$ and $T_C$, we define the upper critical magnetic field ($B_{C2}^{50\%}$) as the field where the resistance is 50% of the normal state resistance, as illustrated by the dashed blue and red arrows in Fig. 5(a). This methodology has been widely used to study $B_{C2}$ in the superconductor research community [19,20]. Clearly, $B_{C2}^{50\%}$ monotonically increases as $V_g$ decreases, as shown in Fig. 5(b), which is opposite to the trend of $T_C$ vs. $V_g$ in the underdoped region. Since $B_{C2}$ is related to the local superconductivity and the coherence length of the Cooper pairs, the opposite trends of $B_{C2}^{50\%}$ and $T_C$ in the underdoped region suggest that the pairing strength of Cooper pairs becomes stronger for lower $n_s$. This feature of $B_{C2}$ increasing with decreasing $n_s$ ($V_g$) has been observed previously in several superconducting systems, including the high temperature cuprates, iron-based superconductors, and the $LaAlO_3/SrTiO_3$ interfaces [15,21,22]. In previous studies, this behavior has been considered as the formation of a "pseudogap" state with local bosonic pairs above $T_C$ [21,23]. The lower $T_C$ with larger pairing strength can be understood as follows: Since the coherence length ($\xi_0$) is shorter in the underdoped region ($\xi_0 = \sqrt{\Phi_0/2\pi B_{C2}}$, where $\Phi_0$ is the quantum flux) [24], the bosonic pairs cannot overlap with each other. Together with the dilution of the bosonic pairs at lower $n_s$, it makes the bosonic pairs harder to condense into a superconducting state, and thus, a lower $T_C$ is observed [22]. Interestingly, the experimental evidences of such "pseudogap" state and the preformation of electron pairs have also been reported in $LaAlO_3/SrTiO_3$ superconducting interfaces [15,25], indicating that there might be a similar pairing mechanism for the superconducting interfaces between the $LaAlO_3/SrTiO_3$ and $KTaO_3$ (111)-based heterostructures.

## V. CONCLUSION

In summary, we report gate tunability of the superconducting state of the $EuO/KTaO_3$ (111) interface and the critical role of $n_s$ in $J_C$ and $B_{C2}$. The highest $T_C$ obtained is ~ 2 K at $n_S \sim 1\times10^{14}$ cm$^{-2}$. The opposite trends of $B_{C2}$ and $T_C$ with respect to $n_s$ suggest a stronger Cooper pairing potential and might be the signature of a "pseudogap" state formed in the underdoped region. Our results could pave the way for investigating the pairing mechanisms of the $KTaO_3$ (111) superconducting interface, and might be helpful for further understanding of superconductivity in low carrier density systems.

## ACKNOWLEDGMENTS

This work is supported by the financial support from National Basic Research Programs of China (No. 2019YFA0308401 and No.2017YFA0303301), National Natural Science Foundation of China (No. 11974025 and 11921005), and the Strategic Priority Research Program of the Chinese Academy of Sciences (No. XDB28000000).

W.Q., Y.M., and J.Y. contributed equally to this work.

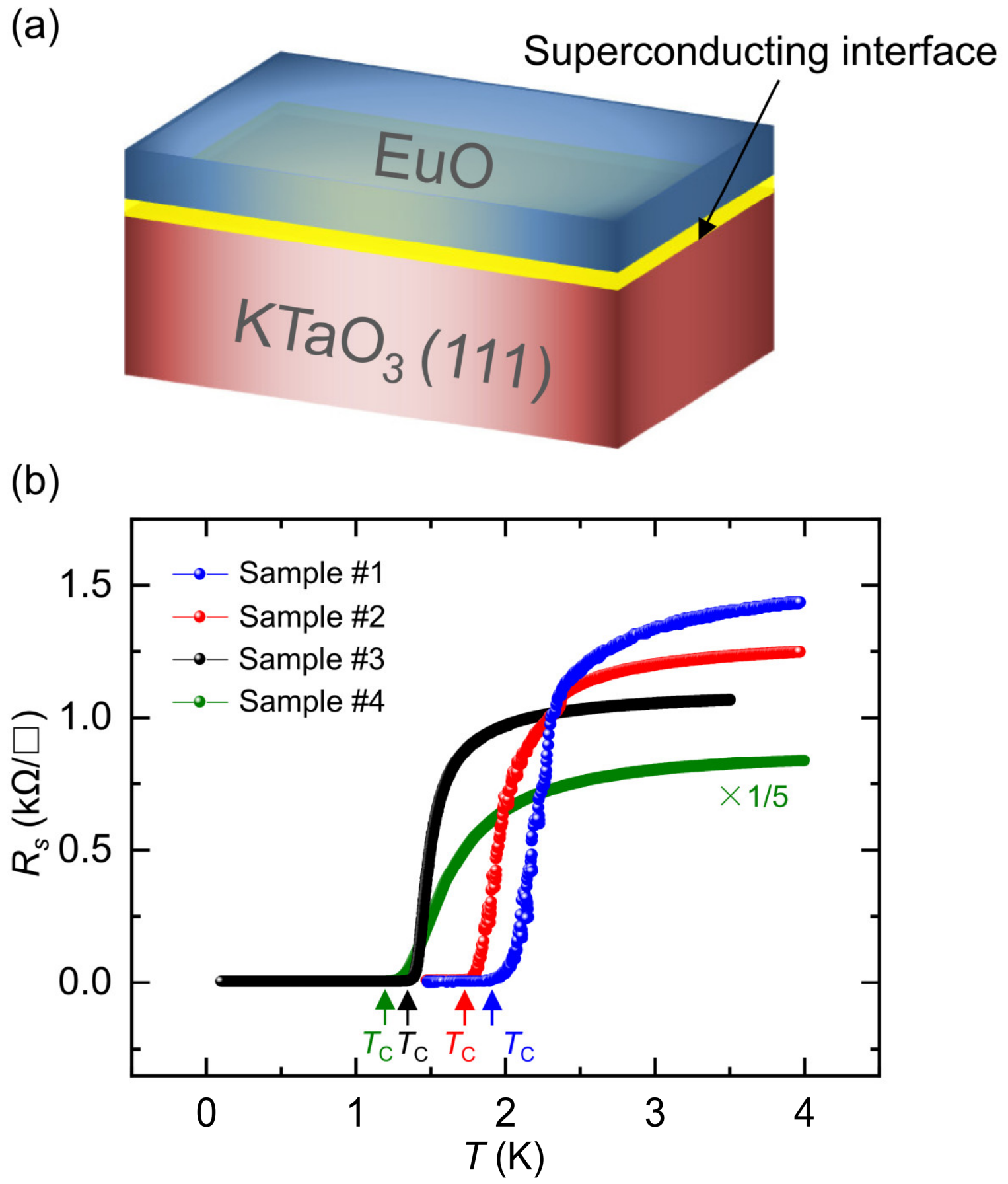

Fig. 1. Schematic and $T_C$ characterization of the superconducting EuO/$KTaO_3$ (111) interfaces. (a) Schematic of the two-dimensional superconducting interface in EuO/$KTaO_3$ (111) heterostructures. (b) Temperature dependence of the sheet resistance ($R_s$) measured on several representative samples with different carrier densities ($n_s$). $T_C$ is defined as the zero-resistance temperature, as indicated by the arrows. $n_s$ of the sample #1 (blue), #2 (red), #3 (black), and #4 (green) are $1.03\times10^{14}$, $9.65\times10^{13}$, $7.97\times10^{13}$ cm$^{-2}$, and $7.74 \times 10^{13}$ cm$^{-2}$, respectively, obtained from Hall measurements at $T = 10$ K.

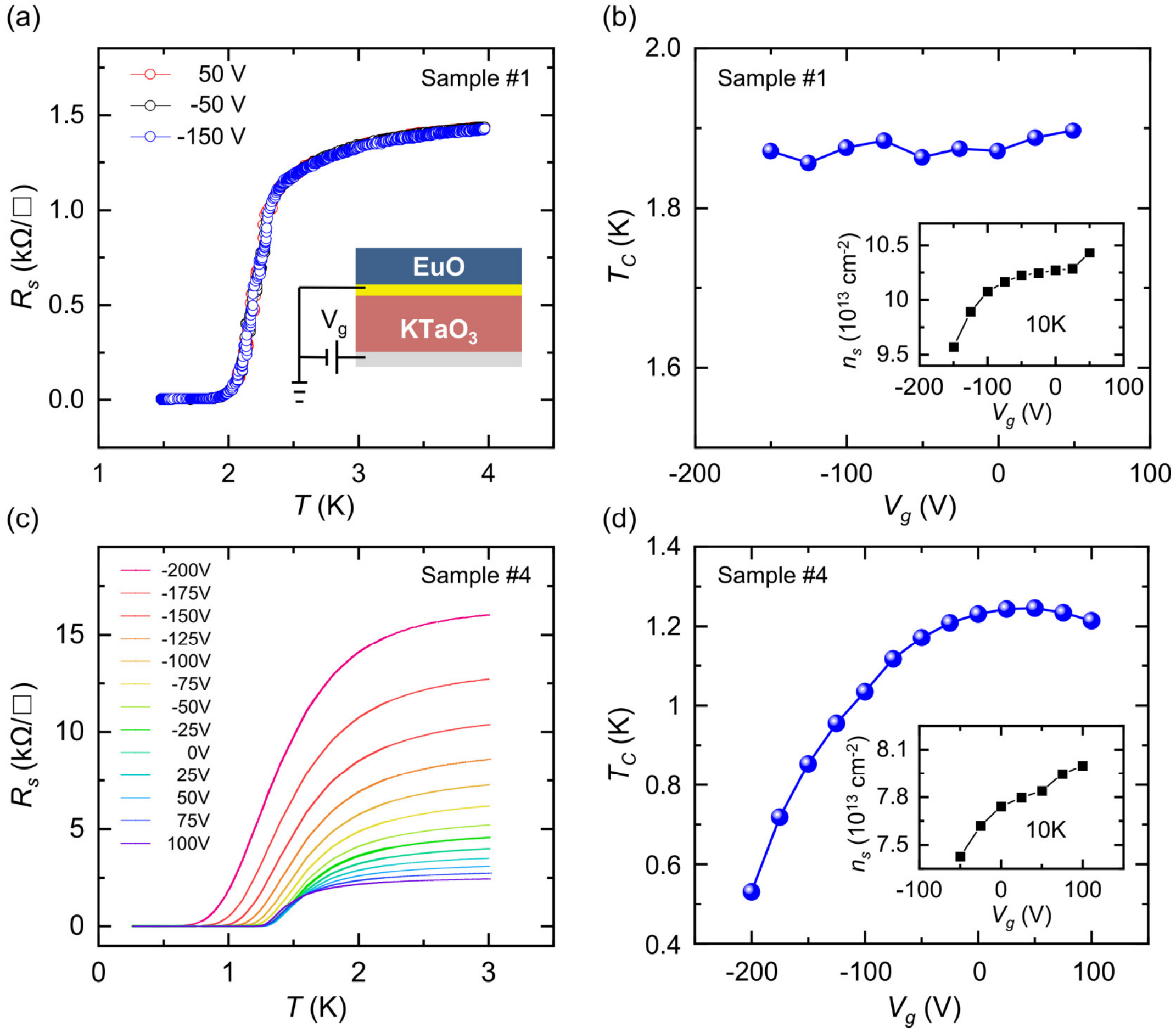

Fig. 2. The gate dependence of $T_C$ of the superconducting EuO/KTaO₃ (111) interfaces. (a) The temperature dependence of $R_s$ of sample #1 at $V_g$ = -150, -50, and 50 V, respectively. Inset: The schematic of the back gate. (b) The gate dependence of $T_C$ of sample #1. Inset: Gate dependence of $n_s$ obtained from Hall measurement at $T$ = 10 K. (c) The temperature dependence of $R_s$ of sample #4 under $V_g$ from -200 to 100 V. (d) The gate voltage dependence of $T_C$ of sample #4. Inset: Gate voltage dependence of $n_s$ obtained from Hall measurement at $T$ = 10 K. For $V_g$ below -50 V, the Hall probe is no longer good for observing linear Hall resistance to determine the actual $n_s$.

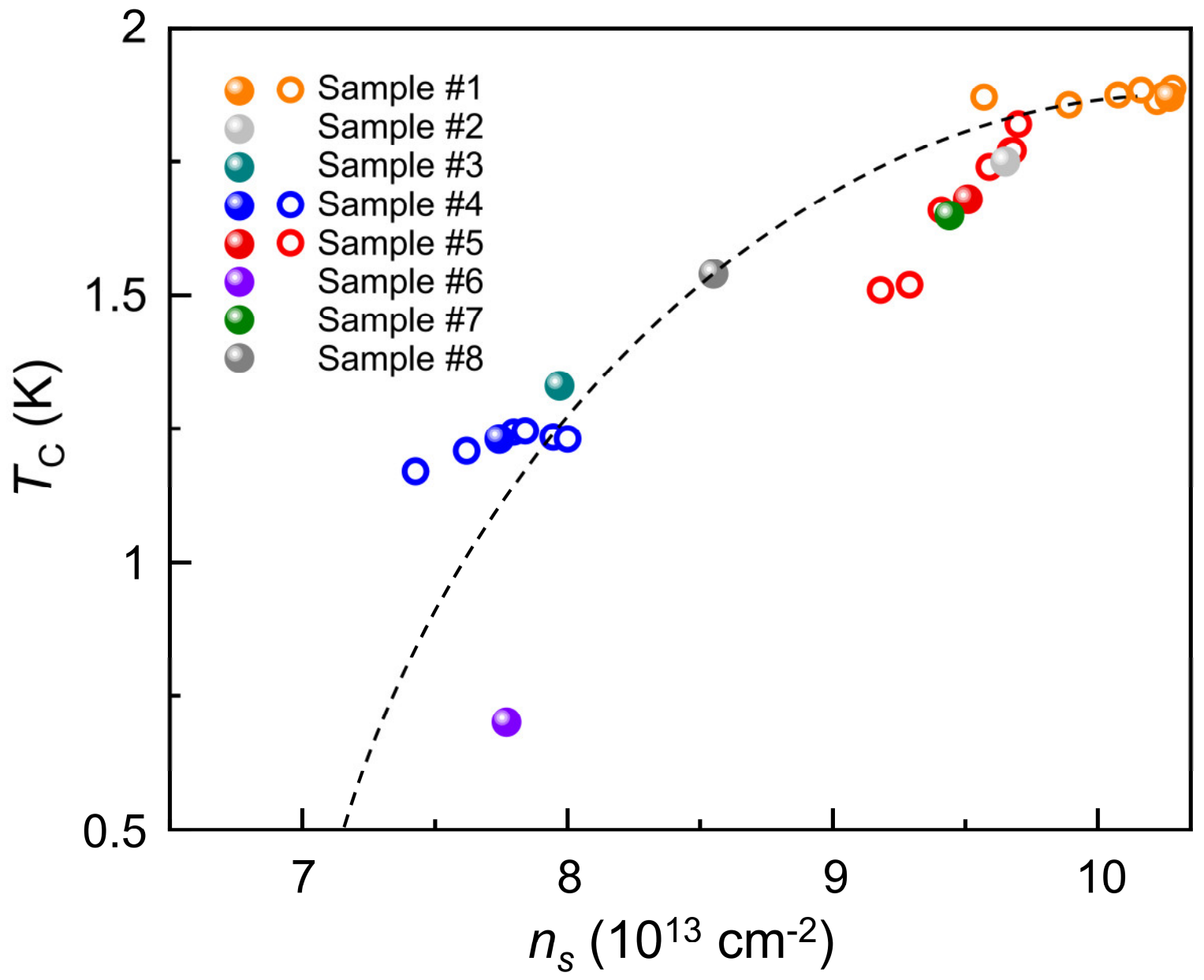

Fig. 3. Superconducting phase diagram of the $EuO/KTaO_3$ (111) interfaces. The solid symbols represent the results obtained from as-grown samples from #1 to #8, and the hollow circles represent the results obtained on the samples (same-color solid balls) tuned by $V_g$. The dashed line is the guide to eye of the boundary of the superconductor phase. Sample #3, #4 and #6 were patterned into Hall device and others were measured using van der Pauw geometry.

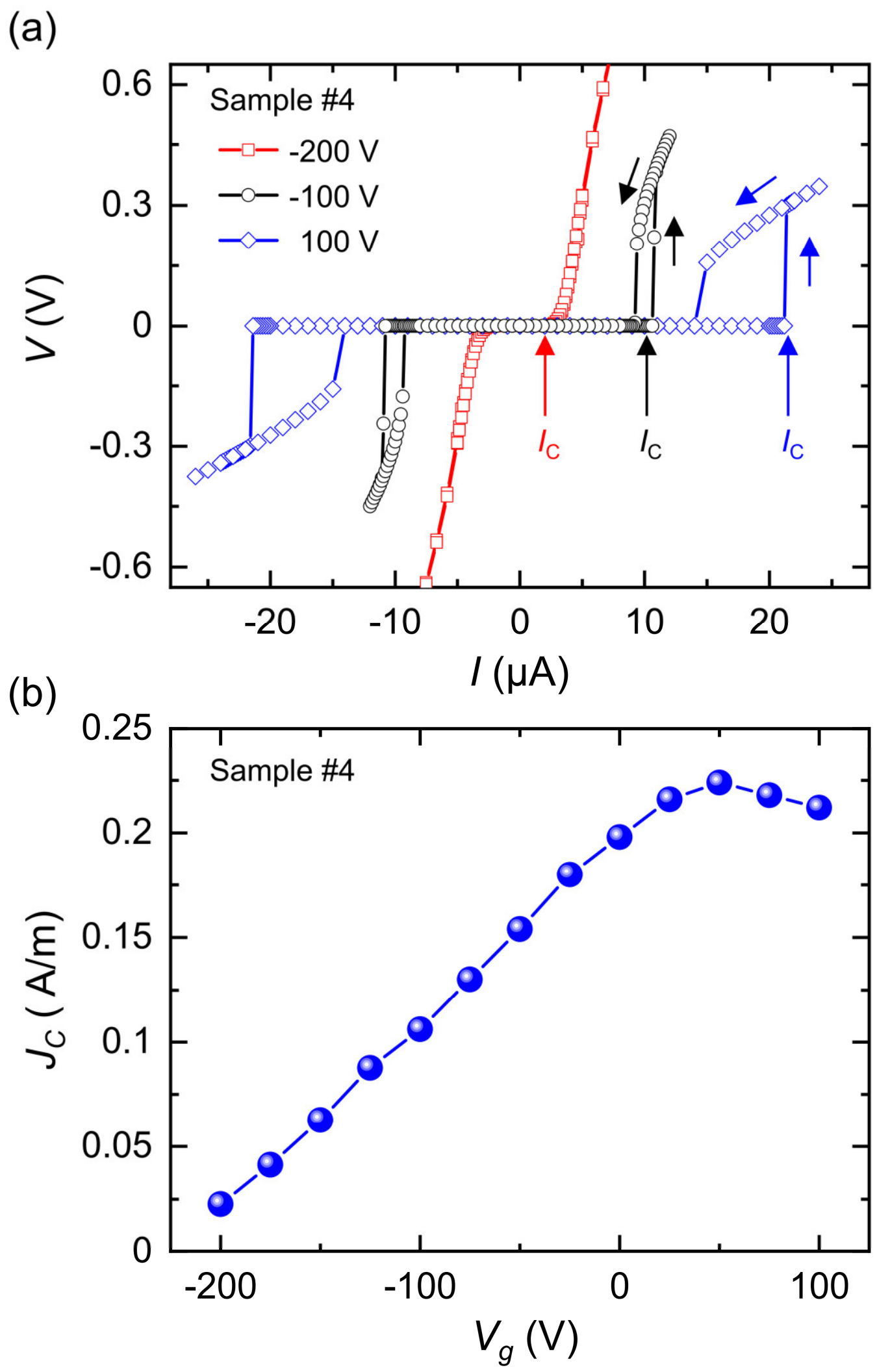

Fig. 4. Gate dependence of the critical current density ($J_C$) of the EuO/KTaO$_3$ (111) interface. (a) The current-voltage (*I*-*V*) curves of the sample #4 under $V_g$ = -200, -100, and 100 V, respectively. The arrows indicate the critical current ($I_C$). (b) The gate voltage dependence of $J_C$ of sample #4 which has a similar trend to the $T_C(V_g)$.

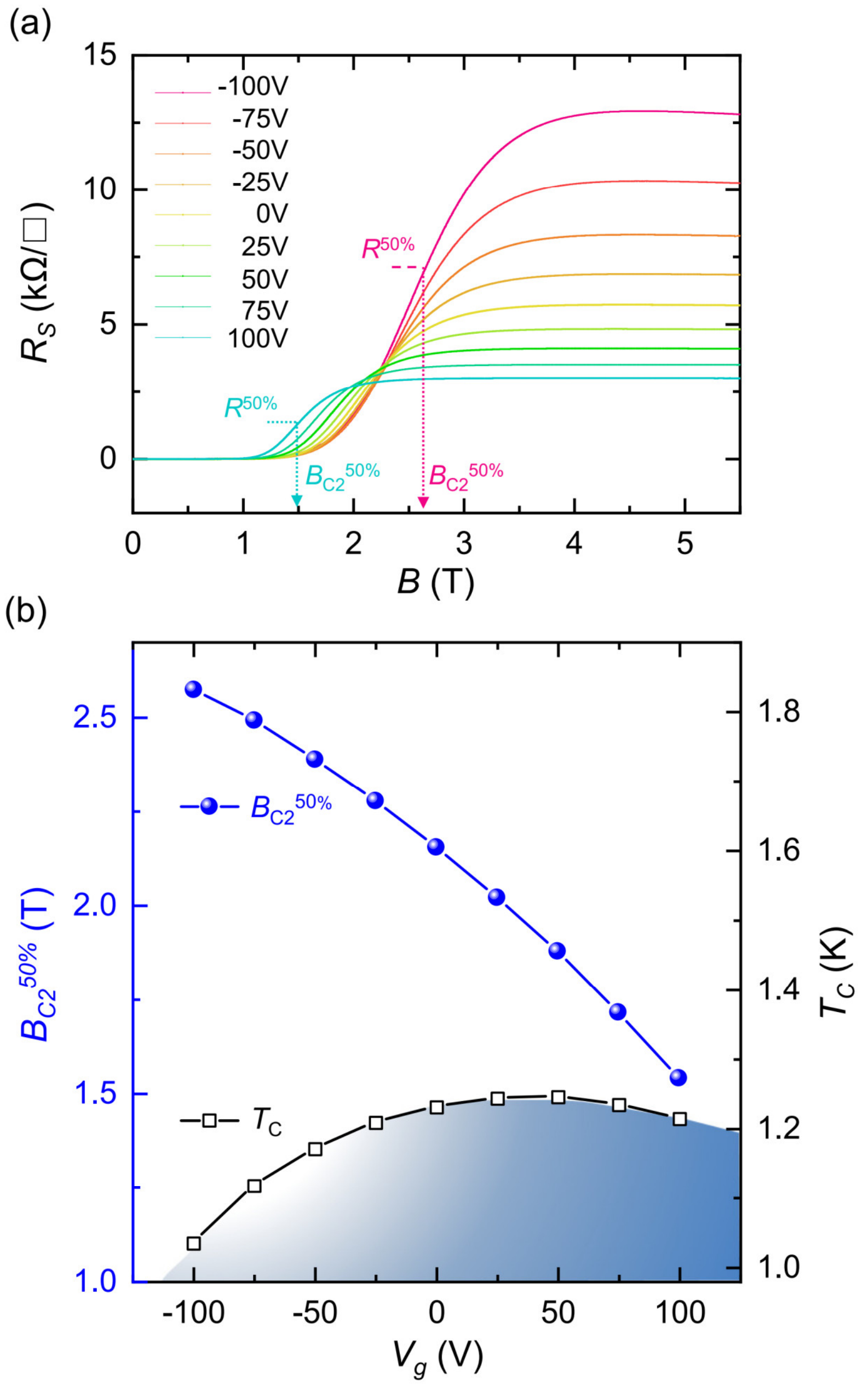

Fig. 5. Gate dependence of the upper critical magnetic field of the EuO/$KTaO_3$ (111) interface. (a) $R_s$ as a function of magnetic field under $V_g$ from -100 to 100 V obtained on sample #4. For the $V_g$ below -100V, the electrical contacts are no longer good to obtain reliable data under magnetic field. The arrows illustrate the determination of $B_{C2}^{50\%}$, under which $R_s$ is half of its normal state value ($R^{50\%}$) at $V_g$ = -100 and 100 V. (b) Gate dependence of $B_{C2}^{50\%}$ and $T_C$.